\documentclass[]{fairmeta}

\title{Bumblebee: Interleaved Mixed-Layer Building Blocks for Large-Scale Recommendation Systems}
\author[1,\dagger]{David Bauer}
\author[1,\dagger]{Cancan Zhang}
\author[1,\dagger]{Wenshun Liu}
\author[1,\dagger]{Xiaoyi Zhang}
\author[1,\dagger]{Weijia Liu}
\author[1,\dagger]{Wanli Ma}
\author[1,\dagger,\ddagger]{Yue Weng}
\author[1]{Wei Li}
\author[1]{Rui Li}
\author[1]{Yiyang Zhao}
\author[1]{Tianqi Lu}
\author[1]{Jing Qian}
\author[1]{Huayu Li}
\author[1]{Xiaoyi Liu}
\author[1]{Linhong Zhu}
\author[1]{Jerry Fu}

\affiliation[1]{Meta}
\contribution[\dagger]{Denotes equal contribution.}
\contribution[\ddagger]{Project lead and corresponding author (\email{yweng@meta.com})}

\abstract{
\db{\textbf{Figures are here: https://fburl.com/4u753m0r}}

\db{\textbf{Eval is here: https://fburl.com/anp/od8i8xfq}}

\db{\textbf{Get industrial dataset size here: N9752932}}

\db{\textbf{Ablation is here: N10332595}}

Recommendation systems have undergone significant transformations in the past years. The transition from traditional feature interaction modules to generative next-action prediction~\citep{zhai2024hstu,ding2026bendingscalinglawcurve} has pushed the boundaries of personalized content. Developments have largely evolved along two separate tracks. Sequence modeling approaches~\citep{zhai2024hstu,pi2020sim} on the one hand and feature interaction methods~\citep{zhang2022dhen,wang2017deepandcross} on the other. In this paper, we introduce \textit{Bumblebee}, a recommendation architecture that addresses the lack of interaction between the two directions through an interleaved, stackable block design. Each block implements a micro-pipeline of layers combining sequence personalization, attention-based encoding, and feature crossing into a self-contained unit. Every block produces a joint representation of both feature modalities which is consumed by the next block in the sequence. This mechanism encourages early and repeated mixture of modalities and enriches downstream features with additional contextual information. Residual connections between blocks create cross-modal information pathways and yield additional predictive performance without adding additional parameters. Blocks can be specialized by selectively dropping components, enabling flexible trade-offs between quality and throughput. We evaluate our approach on large-scale industrial data and show consistent improvements over comparable baseline models across several classification and regression tasks. Furthermore, we conduct ablation studies to confirm that the interleaved composition itself is the primary driver of these improvements. Our results suggest that interleaving heterogeneous functional units, rather than composing deep stacks, is a promising paradigm for future-generation recommendation architectures.

}

\date{\today}

\newcommand{\bbb}{\textit{Bumblebee}}

\definecolor{better}{HTML}{D4EDDA}
\definecolor{bettertext}{HTML}{1E7E34}
\definecolor{worsetext}{HTML}{BD2130}
\definecolor{third}{HTML}{d7e3fc}
\definecolor{second}{HTML}{b7ccf7}
\definecolor{first}{HTML}{99bbff}

\newcolumntype{R}{>{\raggedleft\arraybackslash}X}
\newcolumntype{L}{>{\raggedright\arraybackslash}X}

\newif\ifhasextrabaseline
\hasextrabaselinetrue

\newif\ifhaspublicdatasets
\haspublicdatasetsfalse

\newif\ifshowcomments
\showcommentsfalse

\newif\ifsinglecol
\singlecoltrue

\ifshowcomments
    \newcommand{\db}[1]{\noindent{\textcolor{red}{[DB] #1}}}
    \newcommand{\yw}[1]{\noindent{\textcolor{blue}{[YW] #1}}}
    \newcommand{\wl}[1]{\noindent{\textcolor{brown}{[WL] #1}}}
    \newcommand{\cz}[1]{\noindent{\textcolor{magenta}{[CZ] #1}}}
    \newcommand{\wm}[1]{\noindent{\textcolor{olive}{[WM] #1}}}
\else
    \newcommand{\db}[1]{}
    \newcommand{\yw}[1]{}
    \newcommand{\wl}[1]{}
    \newcommand{\cz}[1]{}
    \newcommand{\wm}[1]{}
\fi

\usepackage{amssymb}
\usepackage{booktabs}
\usepackage{xcolor}
\usepackage{colortbl}
\usepackage{tabularx}
\usepackage{caption}
\usepackage{graphicx}
\usepackage{multirow}
\usepackage{subcaption}
\usepackage{pgfplots}
\usepgfplotslibrary{groupplots}
\usepgfplotslibrary{statistics}
\pgfplotsset{compat=1.18}
\usepackage{bm}

\usepackage{lmodern}
\usetikzlibrary{arrows.meta, positioning, calc, fit, backgrounds}
\tikzset{every picture/.style={/utils/exec={\sffamily}}}

\begin{document}

\maketitle

\db{TODO: add authors/correspondents, code (if available), and blogpost (if available)}

\yw{Yue's comments}

\wl{Wenshun's comments}

\cz{Cancan's comments}

\wm{Wanli's comments}

\section{Introduction}
\label{sec:introduction}

Machine learning-based recommender systems have grown to billions of parameters, making them among the most demanding applications in production machine learning. The adoption of transformer architectures~\citep{vaswani2023attentionisallyouneed}, particularly for sequential user behavior modeling~\citep{zhai2024hstu}, has accelerated this growth by enabling deep attention stacks over user interaction histories (UIH). At the same time, scaling law research has shown that language model performance follows smooth power-law relationships~\citep{kaplan2020scalinglawllm}, and analogous trends have been observed in recommendation systems~\citep{shin2023scalinglawrecommendations}. 


While these findings are encouraging for further architecture development, we observe that modern recommendation architectures have evolved along two largely separate tracks: those focusing on sequence modeling such as HSTU~\citep{zhai2024hstu,ding2026bendingscalinglawcurve}, SIM~\citep{pi2020sim}, and TWIN~\citep{chang2023twin,si2024twinv2}, and more traditional methods focusing on feature interactions like DHEN~\citep{zhang2022dhen}, DCN~\citep{wang2017deepandcross}, and DeepFM~\citep{guo2017deepfm}. While the former are excellent at processing long UIH, contextual data and non-sequence features are often injected before the attention mechanism or near the end of the pipeline. On the other hand, in the latter approaches, sequence features are entirely absent as most such work precedes the recent trend towards generative recommendation. These two tracks have not been deeply integrated. Deep attention stacks often process sequences in isolation from rich non-sequence context or only inject them as additional input, while feature interaction networks consume flattened representations that lose sequential structure. By the time features are combined, important modulation opportunities have been missed.

We introduce \bbb, a recommendation architecture that addresses this blind spot by adopting an interleaved, stackable design that facilitates early and repeated interaction between sequence and non-sequence features. Instead of simply creating deeper transformer blocks to capture user interest, we interleave functional units of generative recommendation and feature interaction modules into compact, self-contained blocks. Each block contains a combination of: (1)~sequence reweighting conditioned on user context, (2)~self attention over UIH, (3)~target-aware filtered attention, (4) cross attention fusing sequence and non-sequence features, and (5)~feature crossing. Since every block produces complete feature representations, we are able to stack multiple blocks for iterative refinement. This design is unlike DHEN~\citep{zhang2022dhen}, which only stacks interaction types, but does not incorporate UIH processing. It also contrasts HSTU~\citep{zhai2024hstu} and similar approaches, which inject non-sequence features either before or after attention. The interleaved design of our method and the resulting modularity also has practical benefits. Blocks are configurable (i.e., dropout of select components), and can be arranged into larger composites to scale performance.


In summary, we make the following contributions:
\begin{itemize}
    \item \textbf{Multi-Stage Feature Interaction.} We propose a novel block architecture that interleaves sequence modeling and feature interaction to capture user behavior at more granular levels. This extends the insight from DHEN~\citep{zhang2022dhen} that heterogeneous interaction modes have complementary strengths.
    \item \textbf{Cross-Module Residual Information Flow.} We show that the interleaved layout fundamentally changes what standard residual connections carry: they implicitly become cross-module pathways that transport both sequence and feature interaction signals. Ablation confirms a $0.25\%$ NE improvement from this mechanism alone, without adding any parameters.
    \item \textbf{Sequence Personalization.} We adapt prior insights from user history personalization~\cite{hou2026kunlunestablishingscalinglaws} to modulate input sequences, giving attention modules a user-centric view on the data.
    \item \textbf{Modular and Composable Architecture.} \bbb~blocks are self-contained units that support configurable component dropout and flexible stacking.
\end{itemize}
\section{Related Work}
This paper is based on prior work on feature interaction modeling, multi-task recommendation, generative user behavior sequence modeling, and the application of large language models to recommendation tasks. In the following paragraphs, we discuss relevant related works from these areas.

\subsection{Feature Interaction and Multi-Task Architectures}

Modeling feature interactions is central to CTR prediction. Modern recommendation systems combine sparse categorical and dense numerical features to produce high-quality predictions. Early architectures separate memorization from generalization. \cite{cheng2016wideanddeep} propose Wide \& Deep networks, combining a wide linear component for memorizing co-occurrences with a DNN for generalization. Following this development,~\cite{wang2017deepandcross} introduce the Deep \& Cross (DCN) architecture, which replaces manual feature engineering with a crossing network to model bounded-degree interactions, while~\cite{guo2017deepfm}'s DeepFM jointly trains a factorization machine for low-order interactions and a deep component for high-order interactions with shared embeddings.

A limitation of these approaches is their uniform interaction type across all feature pairs.~\cite{song2019autoint} address this using multi-head self attention to model high-order interactions, making them more interpretable.~\cite{lang2021aoanforonlinerec} propose AOANet which adaptively selects and combines interaction operations for heterogeneous feature pairs. Many of these works culminate in~\cite{zhang2022dhen}'s DHEN, which makes the observation that different interaction modalities like dot products, self attention, DCNs, and linear models all have non-overlapping strengths across datasets. DHEN is a hierarchical ensemble network that stacks these diverse modes and combines their outputs with residuals. Part of this observation---the stackability of heterogeneous interaction modules---inspired the modular structure of \bbb.

Beyond single-task prediction, industrial problems require the simultaneous optimization of multiple objectives across diverse content domains. This introduces tensions between cross-domain generalization and domain-specific specialization.~\cite{sheng2021onemodeltoservethemall} address this issue for cross-domain scenarios with a star-topology adaptive recommender architecture that can serve multiple domains through a single model by combining a central set of shared parameters with domain-specific parameters via element-wise products. Building on this idea,~\cite{chang2023pepnet} introduce PEPNet with embedding and parameter personalization to serve cross-domain multi-task scenarios using multi-level gating mechanisms.

\subsection{User Behavior Sequence Modeling and Generative Recommendation}

Capturing user interest from long UIH is important for accurate recommendation. This challenge has driven architectural innovation and motivated the shift towards generative formulations. The transformer architecture~\citep{vaswani2023attentionisallyouneed} provided the basic building block of modern sequence modeling. A practical bottleneck is the quadratic memory of standard attention, which~\cite{dao2022flashattention} address with FlashAttention. It is a GPU-architecture-aware tiling algorithm aimed at reducing high-bandwidth IO, making memory complexity linear and enabling attention application on long UIH typical for recommendation tasks.

Early user behavior modeling was limited to short recent histories. However, a growing body of work addresses lifelong sequences by decomposing the problem into efficient retrieval and selective modeling.~\cite{pi2020sim} introduce SIM, which pre-filters UIH based on user-defined matching criteria and then applies attention over the retrieved subset.~\cite{cao2022samplingisallyouneed} extend this idea with locality-sensitive hashing to achieve constant-time retrieval.~\cite{chang2023twin} introduce TWIN to address concerns with SIM, where inadequate pre-filtering conditions lead to performance degradation despite appropriate attention application. Following TWIN, TWIN~V2~\citep{si2024twinv2} extends TWIN's sequence processing capacity from orders of $n=10^4$ to $n=10^5$ through compression and virtual interest representations. More recently,~\cite{chai2025longer} demonstrate scaling laws for sequence length in industrial recommenders. 

The progress in sequence modeling has catalyzed a generative paradigm for recommendation. Rather than scoring individual candidates, generative approaches treat recommendation as a sequence generation problem.~\cite{zhai2024hstu} recognize this trend and introduce HSTU, a sequential transducer model that frames recommendation as next-action prediction. Soon after,~\cite{ding2026bendingscalinglawcurve} extend this line of work with ULTRA-HSTU which replaces compute intensive self attention with more efficient sparse attention mechanisms and optimized sequence design and drastically outscales the original approach.~\cite{deng2025onerec} take a different approach with OneRec, replacing the traditional multi-stage pipeline of retrieval and ranking steps with a single encoder-decoder model. To scale this approach, OneRec uses a sparse Mixture-of-Experts (MoE) of which it uses only a subset during inference.

The underlying paradigm to all these developments is the so-called scaling law.~\cite{kaplan2020scalinglawllm} establish that LLM model performance follows a smooth power-law relationship between model size, data size, and compute. These trends can be observed over several orders of magnitude and indicate that architectural details like deep versus wide have little influence on the overall performance.~\cite{shin2023scalinglawrecommendations} demonstrate analogous scaling laws for recommendation models, showing that user interest representations improve predictably with scale.~\cite{zhang2026zenith} validate these findings with Zenith which introduces an attention-based token fusion approach to unify input features. The important observation they make is that maintaining token heterogeneity is important to avoid collapse and ensure effective scaling. Our design supports this by interleaving attention layers with feature interaction to ensure continuous context injection.
~\cite{hou2026kunlunestablishingscalinglaws} further contribute to better scaling behavior by identifying and optimizing key module bottlenecks commonly found in generative recommenders.

\subsection{Large Language Models and Foundation Models for Recommendation}

The success of foundation models~\cite{touvron2023llama,touvron2023llama2} pre-trained at scale and adapted for downstream tasks has begun to reach and reshape recommendation. Adapting such models to recommendation has been explored along two main axes. One direction uses general foundation models.~\cite{bao2023tallrec} convert recommendation data into natural language tasks and fine-tune a large language model (LLM) to score candidates.~\cite{hou2024llmsarerankers} investigate LLMs as zero-shot rankers, finding that popular models can perform reasonable list-wise ranking when candidates are presented in natural language, although position bias remains challenging. Another direction is to curate specialized models, cotraining them on the same ranking data as downstream applications.~\cite{chen2026massivememorization} propose ViSTA, a foundation model using specialized target attention mechanisms to generate item embeddings from extreme-scale UIH which can be consumed in downstream recommenders.~\cite{lin2025recllmsurvey} provide a survey on use cases of LLMs for recommendation tasks and interested readers may refer to~\cite{pan2025surveyseqrec} for a comprehensive survey on sequential recommendation.

\section{Methodology}

The core of \bbb~is its self-contained block design (Figure~\ref{fig:model_arch}). This design enables easy composition of blocks into larger systems on the one hand. On the other, because each block contains sequence, non-sequence, and feature interaction components, it encourages early multi-modal mixing of feature modalities, which helps personalize inputs for downstream blocks. Furthermore, the design allows for customization of block compositions with configurable dropouts of different components. This flexibility allows for quick adaptation to different problems and input types and further facilitates tailored approaches to system scaling. In the following section, we outline the general composition of a \bbb~block as well as the characteristics of each component. Lastly, we discuss different strategies for block composition and block-wise scaling.

\subsection{Mixed-Layer Blocks}
\begin{figure*}[!htb]
    \centering
    \includegraphics[width=1.0\textwidth]{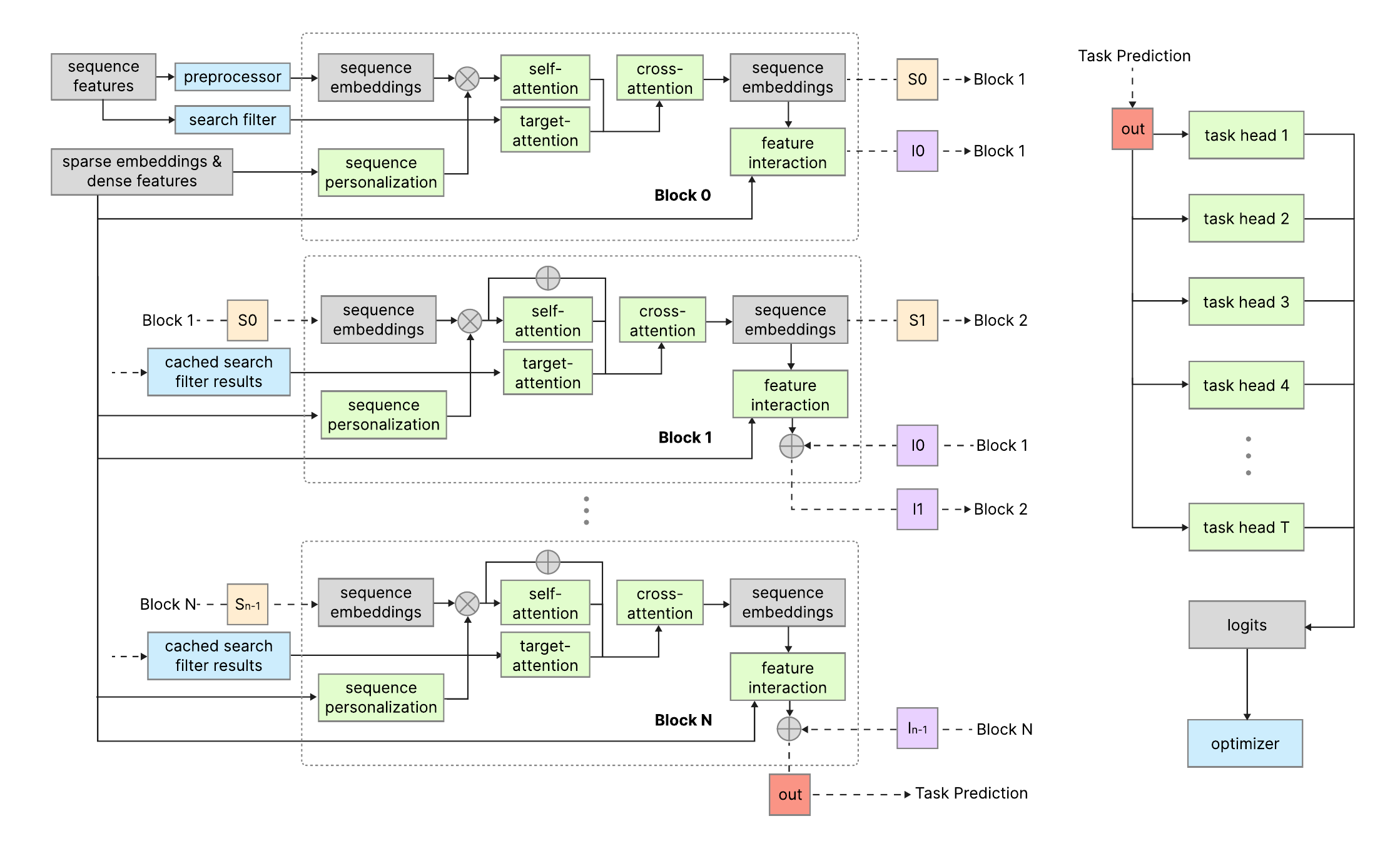}
    \caption{\bbb~model architecture overview. The model consists of multiple self-contained blocks, each containing both sequence and non-sequence modules. Blocks are stacked and connected vertically and residuals create information pathways to share integrated user and candidate context between blocks.}
    \label{fig:model_arch}
\end{figure*}
As discussed in Section~\ref{sec:introduction}, sequence modeling and feature interaction have largely developed as separate paradigms. The \bbb~block is designed to bridge these two tracks in a single, self-contained unit. Each block implements a three-stage micro-pipeline: (1) \textit{personalization}, where non-sequence context modulates raw input sequences; (2) \textit{encoding}, where complementary attention mechanisms extract both global and candidate-specific interest user signals from the personalized sequence; and (3) \textit{integration}, where sequence and non-sequence representations are fused through cross attention and feature crossing. Since every block contains this full pipeline, each block's output is already a joint representation of both feature modalities. This is unlike other architectures~\citep{zhai2024hstu,zhang2022dhen} that defer this fusion to the final layer. 

The general composition of a \bbb~block is outlined in Figure~\ref{fig:model_arch}, where each block constitutes a layer in a vertical stack. As such, the layers contain several components processing different types of feature inputs as shown in Table~\ref{tab:block_components}. In the coming sections, we present each of these components in more detail.

\begin{table}[t]
\caption{Overview of \bbb~block components.}
\label{tab:block_components}
\centering
\small
\begin{tabularx}{\linewidth}{l l X}
\toprule
\textbf{Modularity} & \textbf{Function / Purpose} & \textbf{Description} \\
\midrule
\multirow{1}{*}{\textbf{Personalization}}
& Sequence Reweighting & Reweights interaction history to emphasize relevant signals. \\
\midrule
\multirow{2}{*}{\textbf{Encoding}}
& Global Signal & Self attention over long user interaction history to capture long-term interest. \\
\cmidrule(l){2-3}
& Candidate-Specific Signal & Target-aware interest modeling via pre-filtering long UIH, extracting items relevant to each candidate. \\
\midrule
\multirow{2}{*}{\textbf{Integration}}
& Sequence Enrichment & Augments sequence representations with contextual embeddings from non-sequence features via cross attention. \\
\cmidrule(l){2-3}
& Feature Crossing & Combines non-sequence features and sequence augmentation results, uniting all information derived throughout the block. \\
\bottomrule
\end{tabularx}
\end{table}

\subsection{Sequence Reweighting}
The first stage of each block addresses an important limitation of both paradigms: sequence-focused models process UIH without user-level context, while feature interaction models discard sequential structure completely. Sequence reweighting bridges this gap by conditioning the input sequence on non-sequence features before any attention is applied. 

In sequential recommendation models, each user is associated with their ever-growing history of interactions with items in the system. Individual items, such as social media posts, products, or videos, are generally shared among users in such a way that the same item can appear in many users' histories. However, we make the important observation that the presence of an item in different user interaction histories (UIH) carries vastly different weights and semantic importance among different users and different positions in their histories. That means that two users having identical UIH does not automatically imply identical interests and their ideal next-item recommendation might be different. This is especially relevant in implicit interactions like item viewing duration. Signals like these do not convey explicit intent. This lack of context puts recommendation systems at risk of delivering low-relevance content.

\subsubsection{Reweighting Formulation}
In our design, we address this issue by generating personalized representations of a user's interaction histories. To achieve this, each block contains a \textit{sequence personalization} module (Figure~\ref{fig:model_arch}). This module consumes both non-sequential features such as sparse embeddings and dense features as well as embeddings of the UIH itself (Figure~\ref{fig:uih_sampling} (a)). For a given user $u$ with UIH $S_u = (x_0, x_1,...,x_n)$ $x_i\in\mathbb{X}$, the module predicts sequence reweighting weights $\omega_u$ where $\|\omega_u\| = \|S_u\|$. The sequence input to each block is transformed as follows.
$$
S^{\prime}_{u} = S_u \odot \omega_u
$$
\noindent where $\odot$ is the Hadamard product of UIH items and their respective weights. The reweighted UIH $S^{\prime}_u$ gates each item according to the context derived from the non-sequence features, creating personalized sequence data for downstream modules in each block.

\begin{figure}[t]
\centering
\resizebox{0.85\columnwidth}{!}{%

\definecolor{bgblue}{HTML}{CDEDFD}
\definecolor{bordergray}{HTML}{858585}
\definecolor{goodgreen}{HTML}{7da160}
\definecolor{bgred}{HTML}{FA9485}

\begin{tikzpicture}[
    label/.style={font=\scriptsize\fontseries{l}\selectfont}
]

\node[label, anchor=west] at (0, 0.8) {\textbf{(a) Without Bucketing}};

\node[label, anchor=east] at (-0.2, 0.3) {n = 37};
\draw[fill=bgblue!100, draw=bordergray!100] (0, 0.15) rectangle (0.93, 0.45);

\node[label, anchor=east] at (-0.2, -0.15) {n = 49};
\draw[fill=bgblue!100, draw=bordergray!100] (0, -0.3) rectangle (1.23, 0.0);

\node[label, anchor=east] at (-0.2, -0.6) {n = 89};
\draw[fill=bgblue!100, draw=bordergray!100] (0, -0.75) rectangle (2.23, -0.45);

\node[label] at (0.9, -1.0) {$\vdots$};

\node[label, text=bgred!90!black] at (1.4, -1.5) {O(N) kernels};

\draw[bordergray!30, thick] (3.0, 0.9) -- (3.0, -1.7);

\node[label, anchor=west] at (4.6, 0.8) {\textbf{(b) With Bucketing (ours)}};

\node[label, anchor=east] at (4.4, 0.3) {n = 37};
\draw[fill=bgblue!100, draw=bordergray!100] (4.6, 0.15) rectangle (5.53, 0.45);
\draw[fill=gray!15, draw=bordergray!100, dashed] (5.53, 0.15) rectangle (6.2, 0.45);
\node[label, anchor=west] at (6.3, 0.3) {$\rightarrow$ 64};

\node[label, anchor=east] at (4.4, -0.15) {n = 49};
\draw[fill=bgblue!100, draw=bordergray!100] (4.6, -0.3) rectangle (5.83, 0.0);
\draw[fill=gray!15, draw=bordergray!100, dashed] (5.83, -0.3) rectangle (6.2, 0.0);
\node[label, anchor=west] at (6.3, -0.15) {$\rightarrow$ 64};

\node[label, anchor=east] at (4.4, -0.6) {n = 89};
\draw[fill=bgblue!100, draw=bordergray!100] (4.6, -0.75) rectangle (6.83, -0.45);
\draw[fill=gray!15, draw=bordergray!100, dashed] (6.83, -0.75) rectangle (7.8, -0.45);
\node[label, anchor=west] at (7.9, -0.6) {$\rightarrow$ 128};

\node[label] at (5.5, -1.0) {$\vdots$};

\node[label, text=goodgreen!100] at (5.8, -1.5) {O(10) kernels};

\end{tikzpicture}
}
\caption{Shape bucketing for sequence reweighting. (a)~Without bucketing, each unique sequence length requires a dedicated compiled kernel. (b)~With bucketing, lengths are rounded up to the nearest bucket size, reducing the kernel count to $O(10)$ at the cost of minor padding.}
\label{fig:shape_bucketing}
\end{figure}
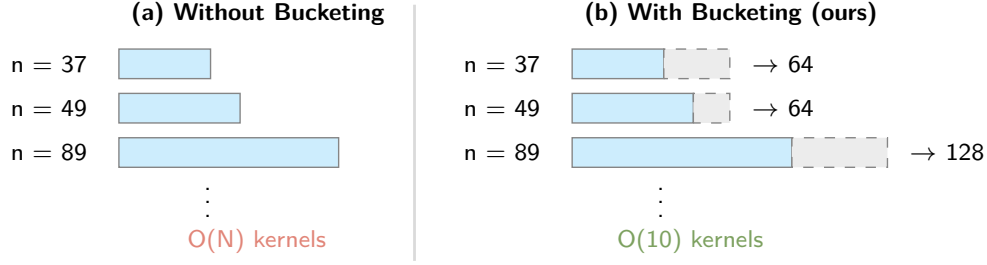

A practical consideration for the reweighting module is computational efficiency. Since UIH lengths vary across users, each unique sequence length triggers a separate kernel compilation under standard eager or compiled execution, potentially leading to thousands of compiled kernels. To mitigate this, we introduce \textit{shape bucketing}: sequence lengths are rounded up to the nearest value in a predefined set of bucket sizes (e.g., 32, 64, 128, \ldots, 4096), reducing the number of distinct compiled kernels to $O(10)$ while introducing only minor padding overhead (Figure~\ref{fig:shape_bucketing}).

\subsubsection{UIH Feature Isolation}
An important design consideration for sequence reweighting is the choice of conditioning features. In industrial recommendation systems, the set of input features naturally divides into \textit{user-specific} features (UF)---attributes that are fixed for a given user's request such as user profile, device type, and time of day---and \textit{candidate-specific} features (CF) that vary per candidate item, such as item category or creator information. Conditioning the reweighting module on the full feature set including UF and CF introduces several problems.

\begin{enumerate}
    \item \textbf{Candidate count mismatch.} During training, each user request is paired with a fixed batch of candidate items, whereas at serving time the number of candidates per request is determined by retrieval and early-stage ranking and typically differs from training. CF features therefore appear with variable multiplicity, creating a distributional gap between training and inference.
    \item \textbf{Candidate distribution mismatch.} Training candidates are post-exposure samples---items that were previously shown to the user and received implicit or explicit feedback. In contrast, candidates that are to be served are produced by upstream retrieval or early-stage ranking and follow a fundamentally different distribution. Conditioning on CF means the reweighting module learns patterns specific to the training distribution that may not transfer to the serving distribution.
    \item \textbf{Score dependency.} If candidate-level CF are aggregated (e.g., averaged) before being passed into the reweighting module, the resulting sequence weights $\omega_u$ become functions of the \emph{entire} candidate set. This makes each candidate's prediction score implicitly dependent on all other candidates in the same request which violates the assumption that item scores should be independent.
\end{enumerate}

\begin{figure}[t]
\centering
\resizebox{0.7\textwidth}{!}{%
    \definecolor{opgreen}{HTML}{E2FDCD}
\definecolor{bordergray}{HTML}{858585}
\definecolor{bgblue}{HTML}{CDEDFD}
\definecolor{bgorange}{HTML}{FACB85}
\definecolor{bggray}{HTML}{D9D9D9}
\definecolor{goodgreen}{HTML}{7da160}
\definecolor{bgred}{HTML}{FA9485}

\begin{tikzpicture}[
    every node/.style={font=\scriptsize},
    box/.style={rectangle, draw, rounded corners=0pt, minimum height=0.4cm, align=center, inner sep=2.5pt},
    robox/.style={box, fill=bgblue!100, draw=bordergray!100},
    nrobox/.style={box, fill=bgorange!100, draw=bordergray!100},
    procbox/.style={box, fill=bggray!100, draw=bordergray!100},
    pffnbox/.style={box, fill=opgreen!100, draw=bordergray!100, minimum width=1.2cm},
    arr/.style={-{Stealth[length=1.2mm]}, semithick},
    xarr/.style={-{Stealth[length=1.2mm]}, semithick, dashed, red!50},
]
\node[font=\scriptsize\bfseries] (ta) at (0, 0) {(a) Without Isolation};

\node[robox, below=0.25cm of ta, xshift=-0.55cm] (ro_a) {UF};
\node[nrobox, right=0.15cm of ro_a] (nro_a) {CF};
\node[procbox, below=0.4cm of $(ro_a)!0.5!(nro_a)$] (mix_a) {All Features};
\node[pffnbox, below=0.35cm of mix_a] (pffn_a) {Sequence Personalization};
\node[below=0.3cm of pffn_a] (out_a) {$\omega_u$};

\draw[arr] (ro_a.south) -- (mix_a);
\draw[arr] (nro_a.south) -- (mix_a);
\draw[arr] (mix_a) -- (pffn_a);
\draw[arr] (pffn_a) -- (out_a);

\node[font=\tiny, text=bgred!90!red, align=center, below=0.15cm of out_a] (warn) {
    \textbf{Train-Serve Distribution Shift}
};

\node[font=\scriptsize\bfseries] (tb) at (4.0, 0) {(b) User Feature Isolation (Ours)};

\node[robox, below=0.25cm of tb, xshift=-0.55cm] (ro_b) {UF};
\node[nrobox, right=0.15cm of ro_b] (nro_b) {CF};
\node[procbox, below=0.4cm of ro_b] (proj_b) {UF Projection};
\node[procbox, below=0.35cm of proj_b] (agg_b) {User-Level Aggregation};
\node[pffnbox, below=0.35cm of agg_b] (pffn_b) {Sequence Personalization};
\node[below=0.3cm of pffn_b] (out_b) {$\omega_u$};

\draw[arr] (ro_b.south) -- (proj_b);
\draw[arr] (proj_b) -- (agg_b);
\draw[arr] (agg_b) -- (pffn_b);
\draw[arr] (pffn_b) -- (out_b);

\draw[xarr] (nro_b.south) -- ++(0, -0.35);
\node[font=\tiny, text=red!60] at ($(nro_b.south)+(0.28,-0.39)$) {\texttimes\ bypass};

\node[font=\tiny, text=goodgreen!90!green, align=center, below=0.15cm of out_b] (ok) {
    \textbf{Train $\boldsymbol{\equiv}$ Serve}
};

\end{tikzpicture}
}
\caption{User-specific feature (UF) isolation for sequence reweighting. (a)~Conditioning the sequence reweighting module on all features including candidate-varying features (CF) leads to a distribution shift between training ($N$ candidates per user) and serving ($M$ candidates). (b)~Our approach restricts sequence reweighting inputs to UF features only, projected through a dedicated layer and aggregated per user, ensuring invariance to candidate count.}
\label{fig:ro_isolation}
\end{figure}
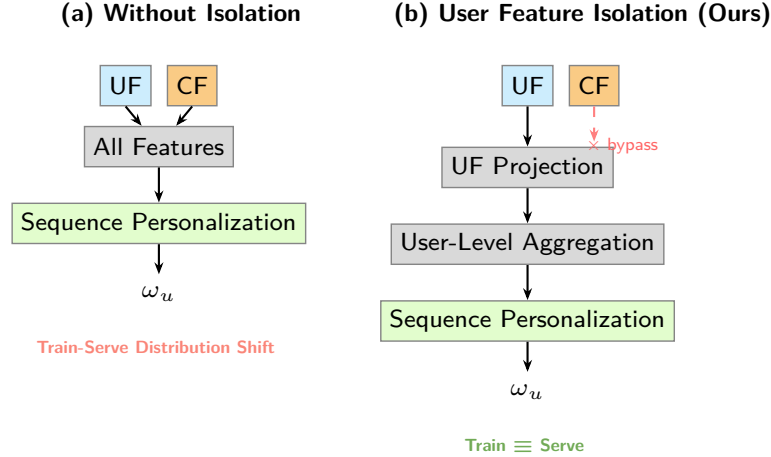

To avoid these issues, we restrict the conditioning input of the sequence reweighting module to UF only. Concretely, we extract sparse embeddings from the set of all embeddings and project UF dense features through a dedicated projection layer, bypassing the shared dense feature path that mixes UF and CF signals. These UF-only representations are aggregated at the user level with one copy per user, independent of candidate count, before being passed into the reweighting module. This ensures that $\omega_u$ depend solely on user-level context and remain invariant to the number and nature of candidates, maintaining consistent behavior between training and serving stages (Figure~\ref{fig:ro_isolation}).

\subsection{Long-Sequence Self Attention}

Having personalized the input sequence, the block next encodes it through two complementary mechanisms. Self attention (this section) captures intra-sequence patterns across the full UIH, while target attention (next section) extracts candidate-specific signals. Together, these two steps produce semantically orthogonal representations where one captures user-centric and the other candidate-centric interest for the integration stage.

For the self attention module, we adopt the HSTU architecture~\citep{zhai2024hstu}, which frames recommendation as next-action prediction over a sequence of user interactions. We refer the reader to~\cite{zhai2024hstu} for architectural details. In the original design, HSTU operates as a standalone deep stack of attention layers processing sequences in isolation from non-sequence context. In \bbb, the role of self attention is fundamentally different in two respects. First, the input sequence has already been personalized by the upstream reweighting module, so attention operates on a user-conditioned representation rather than raw interaction embeddings. Second, the self attention output is not the final representation but an intermediate encoding that feeds into the integration stage, where it is fused with non-sequence features via cross attention and feature crossing. As a result, each block's self attention captures global user interest patterns that are both personalized and integration-ready, rather than serving as the sole feature extractor.

\subsection{Filtered Target Attention}
\label{sec:filtered_attn}
\begin{figure}[!htb]
    \centering
    \includegraphics[width=0.75\columnwidth]{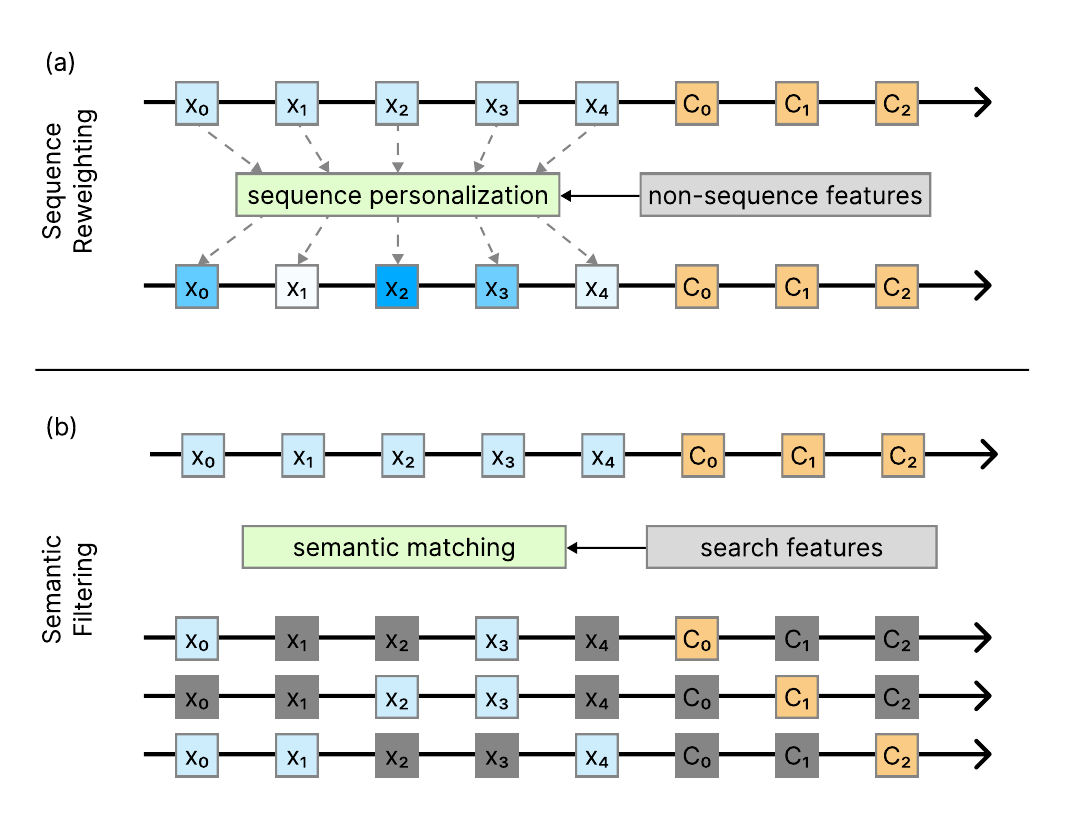}
    \caption{Each block performs two types of sequence resampling. The sequence personalization module re-weights the input sequence to create personalized inputs for each user. The search-based filtering module selects a subset of salient UIH items based on their relevance for each candidate creating candidate-aware embeddings. In combination, these two approaches extract representations of the input sequences covering user and item contexts to create a semantically orthogonal set of features for downstream modules.}
    \label{fig:uih_sampling}
\end{figure}

Where self attention captures users' global interest patterns, filtered target attention addresses the complementary need of identifying which historical interactions are most relevant to a specific candidate item. This candidate-awareness is missing from feature interaction methods and typically isolated in dedicated retrieval stages in sequence-focused architectures. By embedding it within each block alongside self attention and feature crossing, \bbb~ensures that candidate-specific signals inform non-sequence context at every layer. We achieve this by sub-sampling the UIH and extracting relevant entries individually for each candidate item (Figure~\ref{fig:uih_sampling} (b)). This follows the idea of~\cite{pi2020sim}. Unlike prior work, we re-use these subsets to generate multiple embeddings---one in each block. 

To sub-sample the UIH, we define a set of \textit{search features}. These features have corresponding values in UIH entries as well as candidate items. An example of a search feature could be the category of a product (e.g., "household"). Choices for such features are abundant, however we limit our design to categorical features to reduce computational complexity. Candidate-side features are matched against the same feature on UIH entries. Matches are included in the subset and entry collection is stopped after a user-defined maximum number of matches is accumulated. This step is analogous to~\cite{pi2020sim}'s hard-search approach.


\subsection{Cross Attention}
\label{sec:cross_attn}
The integration stage of each block begins with a pooled multi-head cross attention (PMA) module~\citep{lee2019settransformer}, which bridges the sequence and non-sequence tracks. Unlike standard cross attention where one modality directly queries the other, PMA employs a set of $K$ learnable seed embeddings that serve as queries. These seeds are first conditioned on non-sequence features through a linear compression embedding (LCE) layer, which projects the concatenation of sparse embeddings and candidate-level features into the seed space. The conditioned seeds then attend over the encoded user interaction history via multi-head attention, where the sequence embeddings serve as keys and values. Each seed extracts a different aspect of user interest from the sequence, producing $K$ pooled representations that summarize the sequence through the lens of non-sequence context.

Formally, given non-sequence features $\mathbf{F} \in \mathbb{R}^{B \times N_f \times D}$ and sequence embeddings $\mathbf{S} \in \mathbb{R}^{L \times D}$, the PMA module computes:
$$
\mathbf{Q} = \text{LCE}(\mathbf{F}) + \mathbf{E}_{\text{seed}}, \quad \mathbf{O} = \text{MultiHead}(\mathbf{Q}, \mathbf{S}, \mathbf{S})
$$
\noindent where $\mathbf{E}_{\text{seed}} \in \mathbb{R}^{K \times D}$ are learnable seed parameters and $\mathbf{O} \in \mathbb{R}^{B \times K \times D}$ are the pooled cross-modal representations.

This design has two advantages over direct sequence-to-feature cross attention. First, the learnable seeds act as a bottleneck that compresses a variable-length sequence into a fixed-size representation, keeping the output dimension independent of sequence length. Second, conditioning the seeds on non-sequence features allows the attention pattern to be context-aware---different user contexts lead to different query vectors that extract different interest signals from the same history. Because PMA is applied within every block, non-sequence context is injected repeatedly rather than once, allowing deeper blocks to exploit progressively richer cross-modal representations.


\subsection{Feature Interaction}


Following cross attention, the context-enriched sequence embeddings are combined with the full set of non-sequence features through a feature crossing network. In our design, we adopt DHEN~\citep{zhang2022dhen} for this role, which stacks heterogeneous interaction modes (e.g., dot product, self attention, linear) and combines their outputs via residual connections. DHEN is a natural fit for \bbb~because it was designed to exploit the complementary strengths of diverse interaction types---an insight that aligns with our block-level philosophy of mixing heterogeneous functional units.

Importantly, the feature crossing module is a pluggable component in the \bbb~design. While we use DHEN in our experiments, it can be replaced by any feature interaction module without modifying the rest of the block. Alternative choices include DCN~\citep{wang2017deepandcross}, DeepFM~\citep{guo2017deepfm}, or more recent architectures such as Wukong~\citep{zhang2024wukong}. This modularity allows practitioners to select the interaction module that best suits their latency budget, feature characteristics, or domain requirements, further supporting the composable nature of our design.

The output of the feature crossing module serves as both the block's final representation and, via residual connections, the input to the next block. Because this fusion happens within every block rather than once at the end, each subsequent block receives an already-integrated input, enabling progressively richer cross-modal interaction as depth increases.

\subsection{Cross-Module Residual Flow}
\label{sec:residual_flow}
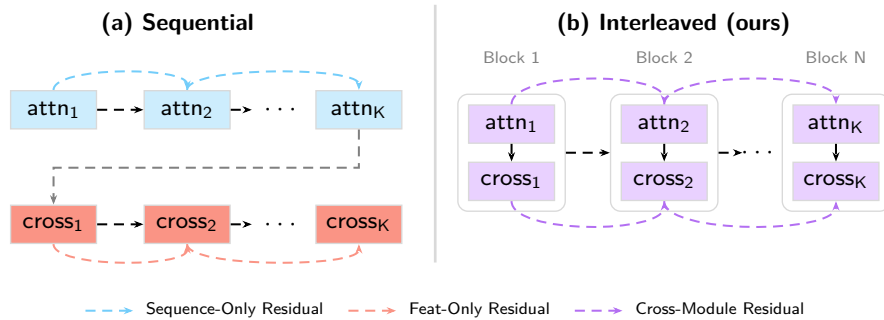
\begin{figure*}[!htb]
\centering
\resizebox{0.80\textwidth}{!}{%
    \definecolor{opgreen}{HTML}{E2FDCD}
\definecolor{bordergray}{HTML}{858585}
\definecolor{bgblue}{HTML}{CDEDFD}
\definecolor{bgorange}{HTML}{FACB85}
\definecolor{bggray}{HTML}{D9D9D9}
\definecolor{bgred}{HTML}{FA9485}
\definecolor{bgpurple}{HTML}{E9D1FF}
\definecolor{fgblue}{HTML}{73cdfa}
\definecolor{fgpurple}{HTML}{b36ff2}

\begin{tikzpicture}[
    every node/.style={font=\scriptsize},
    seqbox/.style={rectangle, draw=bggray!100, fill=bgblue!100, rounded corners=0pt,
        minimum width=0.9cm, minimum height=0.4cm, align=center, inner sep=2pt},
    featbox/.style={rectangle, draw=bggray!100, fill=bgred!100, rounded corners=0pt,
        minimum width=0.9cm, minimum height=0.4cm, align=center, inner sep=2pt},
    mixbox/.style={rectangle, draw=bggray!100, fill=bgpurple, rounded corners=0pt,
        minimum width=0.9cm, minimum height=0.4cm, align=center, inner sep=2pt},
    bblock/.style={rectangle, draw=bggray!100, rounded corners=3pt, inner sep=3pt},
    arr/.style={-{Stealth[length=1mm]}, semithick, densely dashed},
    resblue/.style={-{Stealth[length=1mm]}, fgblue!100, semithick, densely dashed},
    resorange/.style={-{Stealth[length=1mm]}, bgred!100, semithick, densely dashed},
    respurple/.style={-{Stealth[length=1mm]}, fgpurple!100, semithick, densely dashed},
]
\node[font=\scriptsize\bfseries] (ta) at (1.3, 0) {(a) Sequential};

\node[seqbox] (h1) at (0, -0.9) {attn$_\text{1}$};
\node[seqbox] (h2) at (1.4, -0.9) {attn$_\text{2}$};
\node[font=\scriptsize] (hdots) at (2.4, -0.9) {$\cdots$};
\node[seqbox] (hk) at (3.2, -0.9) {attn$_\text{K}$};

\draw[arr] (h1) -- (h2);
\draw[arr] (h2) -- (hdots);

\draw[resblue] (h1.north) .. controls ++(0,0.25) and ++(0,0.25) .. (h2.north);
\draw[resblue] (h2.north) .. controls ++(0,0.25) and ++(0,0.25) .. (hk.north);

\node[featbox] (d1) at (0, -2.1) {cross$_\text{1}$};
\node[featbox] (d2) at (1.4, -2.1) {cross$_\text{2}$};
\node[font=\scriptsize] (ddots) at (2.4, -2.1) {$\cdots$};
\node[featbox] (dk) at (3.2, -2.1) {cross$_\text{K}$};

\draw[arr] (d1) -- (d2);
\draw[arr] (d2) -- (ddots);

\draw[resorange] (d1.south) .. controls ++(0,-0.25) and ++(0,-0.25) .. (d2.south);
\draw[resorange] (d2.south) .. controls ++(0,-0.25) and ++(0,-0.25) .. (dk.south);

\draw[arr, gray] (hk.south) -- ++(0,-0.4) -| (d1.north);

\draw[gray!30, thick] (4.0, 0.2) -- (4.0, -2.5);

\node[font=\scriptsize\bfseries] (tb) at (6.5, 0) {(b) Interleaved (ours)};

\node[font=\tiny, gray] at (4.8, -0.35) {Block 1};
\node[font=\tiny, gray] at (6.4, -0.35) {Block 2};
\node[font=\tiny, gray] at (8.2, -0.35) {Block N};

\node[mixbox] (b1h) at (4.8, -1.05) {attn$_\text{1}$};
\node[mixbox] (b1d) at (4.8, -1.65) {cross$_\text{1}$};
\node[bblock, fit=(b1h)(b1d)] (blk1) {};

\node[mixbox] (b2h) at (6.4, -1.05) {attn$_\text{2}$};
\node[mixbox] (b2d) at (6.4, -1.65) {cross$_\text{2}$};
\node[bblock, fit=(b2h)(b2d)] (blk2) {};

\node[font=\scriptsize] at (7.4, -1.35) {$\cdots$};

\node[mixbox] (bkh) at (8.2, -1.05) {attn$_\text{K}$};
\node[mixbox] (bkd) at (8.2, -1.65) {cross$_\text{K}$};
\node[bblock, fit=(bkh)(bkd)] (blkk) {};

\draw[arr] (b1h) -- (b1d);
\draw[arr] (b2h) -- (b2d);
\draw[arr] (bkh) -- (bkd);

\draw[arr] (blk1.east) -- (blk2.west);
\draw[arr] (blk2.east) -- ++(0.25,0);

\draw[respurple] (b1h.north) .. controls ++(0,0.35) and ++(0,0.35) .. (b2h.north);
\draw[respurple] (b2h.north) .. controls ++(0,0.35) and ++(0,0.35) .. (bkh.north);

\draw[respurple] (b1d.south) .. controls ++(0,-0.35) and ++(0,-0.35) .. (b2d.south);
\draw[respurple] (b2d.south) .. controls ++(0,-0.35) and ++(0,-0.35) .. (bkd.south);

\node[font=\tiny, anchor=west] at (0.2, -3.0) {%
    \tikz[baseline=-0.5ex]{\draw[fgblue!100, semithick, densely dashed, -{Stealth[length=1mm]}] (0,0) -- (0.5,0);}
    \ Sequence-Only Residual
    \quad
    \tikz[baseline=-0.5ex]{\draw[bgred!100, semithick, densely dashed, -{Stealth[length=1mm]}] (0,0) -- (0.5,0);}
    \ Feat-Only Residual
    \quad
    \tikz[baseline=-0.5ex]{\draw[fgpurple!100, semithick, densely dashed, -{Stealth[length=1mm]}] (0,0) -- (0.5,0);}
    \ Cross-Module Residual
};

\end{tikzpicture}
}
\caption{Residual information flow. (a)~In the sequential baseline, attention and interaction
residuals are isolated---they carry only their respective module's information.
(b)~In \bbb's interleaved layout, the same residual mechanisms implicitly
become cross-module pathways: each block's interaction residual already incorporates
sequence encoding, and each block's attention residual receives features refined by
prior interaction steps.}
\label{fig:residual_flow}
\end{figure*}

Between consecutive blocks, residual connections are applied to both the sequence and feature interaction tracks. Specifically, each block's self attention output is connected to its input via a post-layernorm residual, and each block's feature crossing output is connected to the previous block's output through a Squeeze-and-Excitation (SE) scaled residual that performs channel-wise recalibration before addition. These residual forms are standard practice. However, the interleaved layout of \bbb~fundamentally changes the information they carry. In a sequential architecture where all attention layers execute before all interaction layers, attention residuals transport only sequence information and interaction residuals transport only feature crossing signals---the two streams never intersect. In \bbb, because attention and interaction alternate within each block, the feature crossing residual at block $i$ already incorporates the sequence encoding from block $i$'s attention step, and the attention residual at block $i{+}1$ receives a sequence that was personalized using features refined by block $i$'s interaction step. As a result, the same residual mechanism implicitly becomes a cross-module information pathway, allowing deeper blocks to benefit from both earlier sequence patterns and feature interaction signals without requiring any additional architectural components (Figure~\ref{fig:residual_flow}). We empirically validate this design choice in Section~\ref{sec:ablation}: removing cross-module residual connections degrades mean NE by $0.25\%$ despite adding zero parameters, confirming that the gain stems from the information flow enabled by the interleaved layout rather than from additional model capacity.

\subsection{Block Specializations}
\label{sec:block_specializations}
The architecture we presented so far defines a full block, which maximizes cross-modal interaction. However, not all use case scenarios require or are able to afford the full integration. Since each stage is modular, blocks can be specialized by selectively dropping out components and choosing how much of the two-track integration to retain. Enabling such a fine-grained control of each block allows for the construction of mixed-set heterogeneous ensembles as covered in the next section. We validate the contribution of individual components in Section~\ref{sec:ablation}, where the ablation results (Table~\ref{tab:ablation_component}) directly inform the trade-off for each of the components. In the following paragraphs we explore several block specializations for different targeted workflows.

\subsubsection{Full Block}
The simplest configuration is the full block (Figure~\ref{fig:block_specialization} (a)). It exactly corresponds with the modules and flow shown in Figure~\ref{fig:model_arch} and contains a combination of all block components. This can be viewed as a general purpose block that works well in most scenarios but might not be ideal for specialized use cases.

\subsubsection{Lite Block}
A scaled-down version of the full block removes cross and target attention modules to reduce the overall compute cost of the block while retaining some sequence processing capability through the self attention module (Figure~\ref{fig:block_specialization} (b)). This way the block still forms a complete micro-pipeline of functional units while making it more flexible to use in resource-constrained scenarios.

\subsubsection{Long-Sequence Block}
\begin{figure}[!htb]
    \centering
    \includegraphics[width=0.5\linewidth]{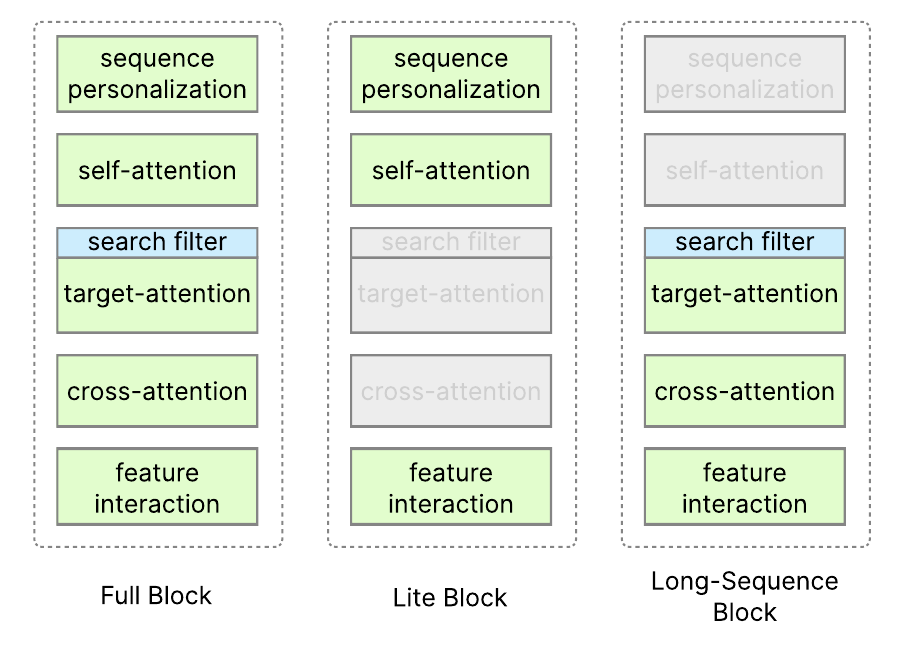}
    \caption{Blocks can be specialized to individual use-cases by dropping out different components that do not fit the use-case. We list three potential configurations that we used throughout our evaluation. (a) Full Block: All components are enabled. Suited for general purpose applications. (b) Lite Block: Drops out additional attention modules to improve runtime performance. Suitable for building lighter models or when creating composites of more than 10 blocks. (c) Long-Sequence Block: Drops self attention and the personalization module and keeps only the filtered target attention and cross attention. Suitable for long-sequence inputs.}
    \label{fig:block_specialization}
\end{figure}

Including a mix of target, cross, and self attention over the full UIH can have considerable impact on model runtime. Self attention scales $O(n^2)$, while others also exhibit high compute cost in large serving scenarios with hundreds or thousands of candidate items. As such, prior approaches~\citep{zhai2024hstu} tend to apply these types of attention on a truncated version of the UIH that covers the $n$ most recent interactions. However, previous work also shows the value of tending to longer sequences~\citep{pi2020sim}. To allow such processing in \bbb, we propose a long-sequence specialization block (Figure~\ref{fig:block_specialization} (c)), that drops out self attention mechanisms while keeping filtered and cross attention (Sections~\ref{sec:filtered_attn},~\ref{sec:cross_attn}) to process fixed-length subsets of much longer input sequences while also retaining a light-weight cross attention module. These longer sequences are used directly by this block type before they are truncated for other blocks.



\section{Evaluation}

We evaluate \bbb~on large-scale industrial data.
To study our design's performance we compare it against \ifhasextrabaseline three production baseline models \else a production baseline model \fi trained on the same data and infrastructure to isolate the architectural contributions.

\subsection{Experiment Conditions}
Our evaluation is split into two themes. First, we compare general model performance against state-of-the-art baselines to evaluate the quality of recommendations derived from each architecture. To this end, we stick closely to the original configuration of each model. Next, we test and ablate our model design, comparing it against a carefully configured baseline containing the \textit{same} components as our design but in a sequential fashion, mimicking the architecture of most current models. While the first part positions \bbb~among comparable approaches, the second part highlights the specific architectural innovation of our interleaved design.

\subsubsection{Models}
As mentioned above, we consider two sets of baselines. On the one hand, a number of state-of-the-art models that we adopt from their respective source publications and adapt to our use case, and on the other hand, a baseline based on modules from our design that we arrange in a sequential fashion, similar to state-of-the-art models, and for which we tune the number of parameters to match our design for a fair comparison.

\textbf{State-of-the-Art Baselines.}
For the general study we choose a DLRM-style model~\citep{dlrm}, HSTU~\citep{zhai2024hstu}, and SIM~\citep{pi2020sim}. The rationale for choosing these models as baselines---aside from being representative state-of-the-art models---is that each of these models represents a core idea of the functional units in each \bbb~block, such as self attention, long-sequence processing, and feature interaction.

\textbf{Equal-Parameter Baseline.}
For the parameter-controlled study, we compare our method against a baseline model, which we call \textit{Sequential}, using the same functional components as our model. Instead of creating multiple blocks and interleaving functions, we combine layers of the same kind together into deep stacks and arrange them sequentially which mirrors the design of most models. The model is represented as a single ``megablock'' with the aforementioned deep stacks of attention and interaction layers. To ensure fairness, we configure both models with broadly the same number of parameters and ensure the overall FLOPs per sample match between models (Table~\ref{tab:model_params}).

\begin{table}[!htb]
    \centering
    \caption{Model parameter counts for the equal-parameter study.}
    \smallskip
    \small
    \begin{tabularx}{0.55\linewidth}{L R R}
    \toprule
    \textbf{Component} & \textbf{Sequential} & \textbf{Bumblebee (ours)} \\
    \midrule
    Sparse Parameters     & 3.802T & 3.802T \\
    Dense Parameters      & 1.914B & 1.952B \\
    \midrule
    Total Parameters      & 3.804T & 3.804T \\
    \bottomrule
    \end{tabularx}
    \label{tab:model_params}
\end{table}

While both models share a similar number of parameters, their compositions differ notably. The baseline model is arranged as a ``megablock'', modeling the structure of current transformer-based recommendation models. This means that these models are made up of a single block, stacking any multi-tier components like attention directly onto each other. This is unlike \bbb, which separates and interleaves deep module stacks. For our baseline, we combine one self attention layer, four layers of target attention, and four layers of cross attention followed by four layers of feature interaction. For the \bbb~condition, we combine a set of heterogeneous blocks (Section~\ref{sec:block_specializations}). Namely, we use one full block (Figure~\ref{fig:block_specialization} Left) followed by three long-sequence blocks (Figure~\ref{fig:block_specialization} Right).


\subsubsection{Datasets}

\ifhaspublicdatasets\else We train all models on \fi a large-scale industrial production dataset (Table~\ref{tab:dataset_metadata}). This rich collection of data is representative of industry-scale recommendation systems and serves as a yardstick for all of our metric results.

\begin{table}[!htb]
    \centering
    \caption{Datasets used for evaluation. The counts for the industrial dataset reflect the amount of training data used to train each of the models in our evaluation. User and Item counts are unique counts per 1-hour timestep (TS). We train all models on 72 TS in total.}
    \smallskip
    \small
    \begin{tabularx}{0.7\linewidth}{L R R R R}
    \toprule
    \textbf{Dataset} & \textbf{Unique Users per TS} & \textbf{Unique
    Items per TS} & \textbf{\#TS} & \textbf{Total \#Samples}  \\
    \midrule
    Industrial & 155.2M & 50.8M & 72 & 81B \\ 
    \bottomrule
    \end{tabularx}
    \label{tab:dataset_metadata}
\end{table}

\subsection{Model Performance}

\begin{figure*}[!htb]
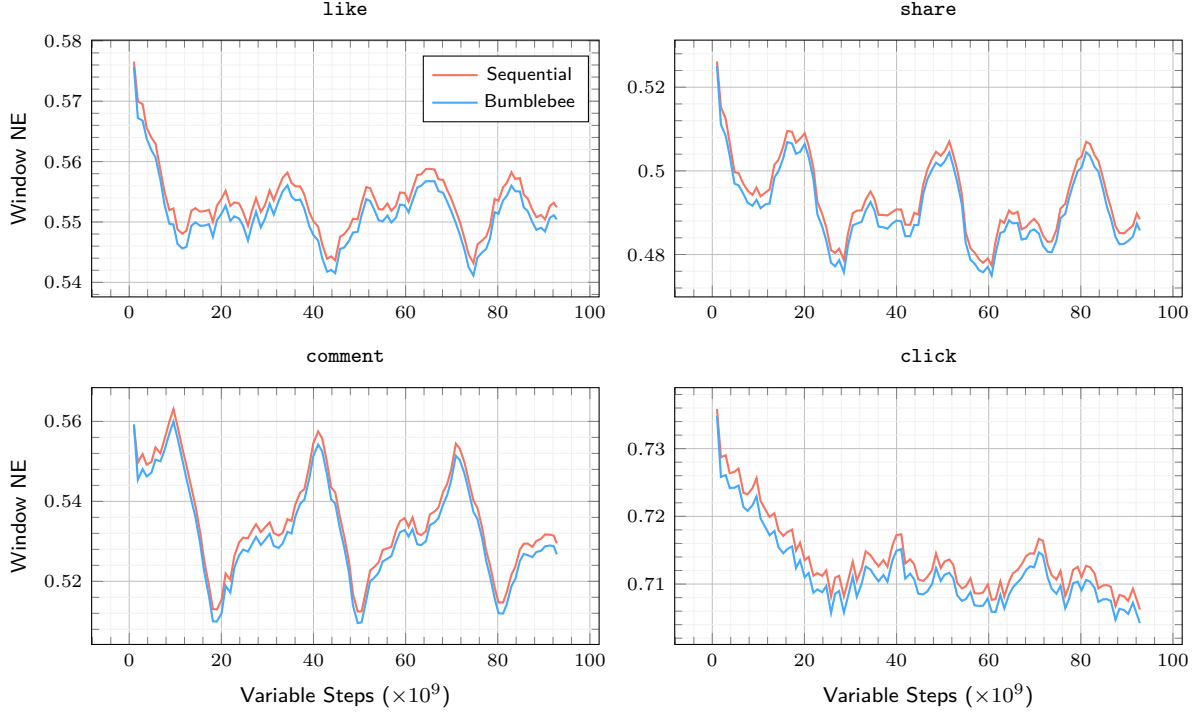

\centering
\begin{tikzpicture}
\begin{groupplot}[
  group style={group size=2 by 2, horizontal sep=1.0cm, vertical sep=1.2cm},
  width=0.50\textwidth,
  height=0.30\textwidth,
  legend style={font=\scriptsize, at={(0.98,0.94)}, anchor=north east},
  grid=both,
  grid style={line width=.1pt, draw=gray!10},
  major grid style={line width=.2pt,draw=gray!50},
  minor tick num=4,
  tick label style={font=\scriptsize},
  label style={font=\footnotesize},
  title style={font=\footnotesize},
  restrict x to domain=1:100,
]
\nextgroupplot[title={\texttt{like}}, ylabel={Window NE}]
\input{figures/pgfplots-ne-like-baseline.tex}
\input{figures/pgfplots-ne-like-candidate.tex}
\legend{Sequential, Bumblebee}

\nextgroupplot[title={\texttt{share}}]
\input{figures/pgfplots-ne-share-baseline.tex}
\input{figures/pgfplots-ne-share-candidate.tex}

\nextgroupplot[title={\texttt{comment}}, ylabel={Window NE}, xlabel={Variable Steps ($\times 10^9$)}]
\input{figures/pgfplots-ne-comment-baseline.tex}
\input{figures/pgfplots-ne-comment-candidate.tex}

\nextgroupplot[title={\texttt{click}}, xlabel={Variable Steps ($\times 10^9$)}]
\input{figures/pgfplots-ne-click-baseline.tex}
\input{figures/pgfplots-ne-click-candidate.tex}
\end{groupplot}
\end{tikzpicture}
\caption{Window NE convergence over training for four key engagement tasks comparing two equal-parameter models. \bbb~(red) consistently achieves lower NE than \textit{Sequential} (blue), with the gap widening as training progresses.}
\label{fig:ne_convergence}
\end{figure*}

We train \ifhasextrabaseline all the baselines \else the baseline \fi and \bbb~on identical hardware (B200, 8$\times$8 GPUs).
All prediction metrics are computed on a held-out evaluation window and all throughput and latency readings are from the longer training window. For all models we show the average normalized entropy (NE) (Table~\ref{tab:ne_comparison}) and RMSE (Table~\ref{tab:rmse_comparison}) across 26 and four tasks respectively. Furthermore, we show windowed metric throughout model training for several key tasks for the equal-parameter models in Figure~\ref{fig:ne_convergence}. Results show that \bbb~consistently outperforms the \ifhasextrabaseline baselines \else baseline \fi in all consumption and engagement tasks.

\begin{table}[!htb]                       
    \centering
    \caption{Aggregate normalized entropy (26 tasks, lower is better) and RMSE (4 regression tasks, lower is better) results compared to baseline results. NE is divided into \textit{c}, consumption tasks, and \textit{e}, engagement tasks.}
    \small                                 
    \begin{minipage}[t]{0.75\linewidth}       
        \centering                                
        \subcaption{Task Performance, Normalized Entropy}
        \begin{tabular}{l c c c c c c}
        \toprule
        \textbf{Model} & \textbf{c-NE} & $\bm{\sigma^2_c}$ & $\Delta_{NE}$ & \textbf{e-NE} & $\bm{\sigma^2_e}$ & $\Delta_{NE}$\\
        \midrule
        \ifhasextrabaseline
        DLRM  & 0.8034 & $8.07\cdot10^{-4}$ & -- & 0.6358 & $1.15\cdot10^{-2}$ & --\\
        \fi
        \ifhasextrabaseline
        SIM  & 0.7968 & $7.02\cdot10^{-4}$ & $-0.66\%$ & 0.6312 & $1.16\cdot10^{-2}$ & $-0.460\%$\\
        \fi
        HSTU & 0.7964 & $8.01\cdot10^{-4}$ & $-0.69\%$ & 0.6318 & $1.17\cdot10^{-2}$ & $-0.400\%$  \\
        BumbleBee (ours) & \cellcolor{better}\textbf{0.7895} & $7.22\cdot 10^{-4}$ & $-1.387\%$ & \cellcolor{better}\textbf{0.6258} & $1.17\cdot10^{-2}$ & $-1.002\%$ \\
        \bottomrule
        \end{tabular}
        \label{tab:ne_comparison}
    \end{minipage}%
    \hfill
    \\
    \vspace{2em}
    \begin{minipage}[t]{0.40\linewidth}
        \centering
        \subcaption{Task Performance, RMSE}
        \begin{tabular}{l c c c}
        \toprule
        \textbf{Model} & \textbf{c-RMSE} & $\bm{\sigma^2_c}$ & $\Delta{NE}$\\
        \midrule
        \ifhasextrabaseline
            DLRM  & 0.9446 & $0.220$ & --\\
        \fi
        \ifhasextrabaseline
            SIM  & 0.9408 & $0.219$ & $-0.363\%$\\
        \fi
            HSTU & 0.9380 & $0.217$ & $-0.645\%$  \\
            BumbleBee (ours) & \cellcolor{better}\textbf{0.9344} & $0.216$ & $-1.005\%$ \\
        \bottomrule
        \end{tabular}
        \label{tab:rmse_comparison}
    \end{minipage}
\end{table}

\subsection{Training Throughput}

\begin{figure}[!htb]
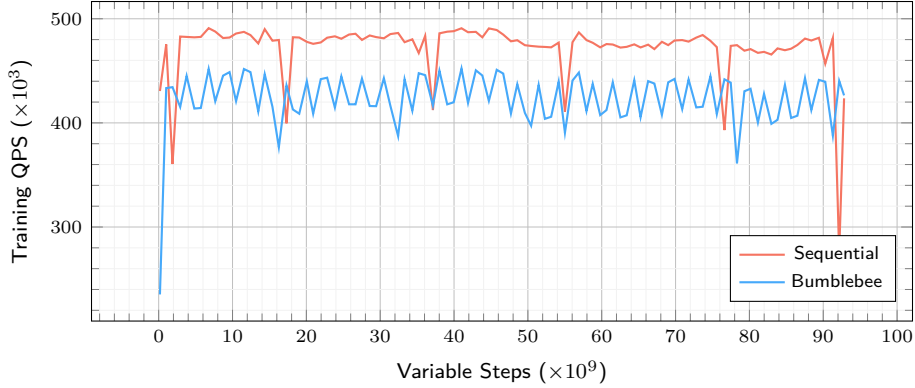

    \centering
    \begin{tikzpicture}
    \begin{axis}[
        width=0.75\textwidth,
        height=0.35\textwidth,
        xlabel={Variable Steps ($\times 10^{9}$)},
        ylabel={Training QPS ($\times 10^{3}$)},
        legend style={font=\scriptsize, at={(0.98,0.06)}, anchor=south east},
        grid=both,
        grid style={line width=.1pt, draw=gray!10},
        major grid style={line width=.2pt,draw=gray!50},
        minor tick num=4,
        tick label style={font=\scriptsize},
        label style={font=\footnotesize},
        title style={font=\footnotesize},
    ]
    \input{figures/pgfplots-qps-baseline.tex}
    \input{figures/pgfplots-qps-candidate.tex}
    \legend{Sequential, Bumblebee}
    \end{axis}
    \end{tikzpicture}
    \caption{Training throughput in queries per second (QPS) over training for the \textit{Sequential} baseline vs.\ \bbb~on B200 hardware.  \bbb's lower QPS reflects its overall complexity; both complete training in comparable wall-clock time.}
    \label{fig:qps_training}
\end{figure}

We compare the training throughput of \bbb~against the \textit{Sequential} baseline on identical hardware using 64 NVIDIA B200 GPUs (Figure~\ref{fig:qps_training}). Over time our model is roughly $7$--$10\%$ slower than a comparable ``megablock'' baseline with the same number of parameters (Table~\ref{tab:model_params}). We attribute this delta to the differences in memory access patterns and operator composition between the two models and because stacked attention blocks generally lend themselves better to kernel fusion. This limitation also hints at interesting future work to improve runtime performance through specialized fusion of heterogeneous components that succeed each other in the \bbb~architecture.

\subsection{Ablation Study}
\label{sec:ablation}

We ablate the key design choices of \bbb~to understand the contribution of each component.


\subsubsection{Component Ablation}
We ablate layer configuration by selectively removing individual components to understand and justify the contribution of different block elements within a \bbb~layer. To this end, we create a baseline configuration containing all tested components and six ablated models with key components removed. Specifically, we remove sequence reweighting, self attention, target attention, cross attention, feature crossing, and the block residual flow. Results are shown in Table~\ref{tab:ablation_component}.

Feature crossing (DHEN) contributes the most ($+0.36\%$ NE), followed by cross attention ($+0.31\%$), and cross-module residual flow ($+0.25\%$). These three components all belong to the integration stage of the block, confirming that the early and repeated fusion of sequence and non-sequence features is the primary driver of \bbb's performance. Sequence reweighting and target attention each contribute $+0.15\%$ which supports the inclusion of the personalization and candidate-aware encoding stages. Self attention shows the smallest individual impact ($+0.05\%$).

\begin{table}[!htb]
\centering
\caption{Component ablation on industrial data. NE is averaged across tasks (lower is better). $\Delta$ is relative to the full model. QPS and memory figures are subject to minor variance due to preemption and resource contention on shared training infrastructure. Colors encode the top three regressions per metric.}
\smallskip
\small
\begin{tabularx}{0.75\linewidth}{>{\raggedright\arraybackslash}p{4.2cm} R R R R}
\toprule
\textbf{Configuration} & \textbf{Mean NE} & \textbf{$\Delta_{\text{NE}}$} & \textbf{$\Delta_{\text{QPS}}$} & \textbf{$\Delta_{\text{Mem}}$} \\
\midrule
BumbleBee (full)           & 0.5958 & -- & -- & -- \\
w/o Sequence Reweighting   & 0.5967 & $+0.15\%$ & \cellcolor{second}$+2.7\%$ & \cellcolor{third}$-1.1\%$ \\
w/o Self Attention         & 0.5961 & $+0.05\%$ & \cellcolor{first}$-0.4\%$ & \cellcolor{second}$+4.7\%$ \\
w/o Target Attention       & 0.5967 & $+0.15\%$ & $+5.2\%$ & $\pm0.0\%$ \\
w/o Cross Attention        & 0.5976 & \cellcolor{second}$+0.31\%$ & $+15.5\%$ & $-3.0\%$ \\
w/o Feature Crossing       & 0.5979 & \cellcolor{first}$+0.36\%$ & $+22.6\%$ & $-5.0\%$ \\
w/o Residual Flow          & 0.5973 & \cellcolor{third}$+0.25\%$ & \cellcolor{third}$+3.4\%$ & \cellcolor{first}$+5.0\%$ \\
\bottomrule
\end{tabularx}
\label{tab:ablation_component}
\end{table}


\subsubsection{Composition Ablation}

To isolate the effect of interleaving from the choice of components, we compare \bbb~against the \textit{Sequential} baseline that contains the \emph{same} functional components---self attention, target attention, sequence reweighting, and feature crossing---but arranges them in a sequential pipeline: all attention layers execute first, followed by all feature interaction layers. This setup ensures that any performance difference is attributable to \emph{how} the components are arranged, not \emph{which} components are used.

As shown in Table~\ref{tab:ablation_composition}, the interleaved composition achieves a $0.20\%$ NE improvement over the sequential arrangement. This confirms that the benefit of \bbb~is not simply from combining more component types, but from the repeated alternation of sequence encoding and feature interaction within each block, which enables the cross-module residual pathways described in Section~\ref{sec:residual_flow}.
\begin{table}[!htb]
\centering
    \caption{Composition ablation: sequential arrangement vs.\ \bbb's interleaved arrangement using the same components.}
\smallskip
\small
\begin{tabularx}{0.85\linewidth}{>{\raggedright\arraybackslash}p{5cm} R R R R}
\toprule
\textbf{Configuration} & \textbf{Mean NE} & \textbf{$\Delta_{\text{NE}}$} & \textbf{$\Delta_{\text{QPS}}$} & \textbf{$\Delta_{\text{Mem}}$}\\
\midrule
Sequential (same components)  & 0.5970 & -- & -- & -- \\
BumbleBee (interleaved)       & \cellcolor{better}\textbf{0.5958} & $-0.20\%$ & $-4.2\%$ & $+3.5\%$ \\
\bottomrule
\end{tabularx}
\label{tab:ablation_composition}
\end{table}




\section{Discussion}
\label{sec:discussion}
Our evaluation shows several interesting findings that inform the design of \bbb's mixed-layer architecture. In the following sections, we discuss the most notable insights from our study and their implications.

\subsection{Integration Over Encoding}
The component ablation in Table~\ref{tab:ablation_component} shows a clear ranking among block modules. The three integration components together account for a large share of the performance of the model than the other components. Namely, feature crossing, cross-attention, and the cross-module residual flow make for $+0.36\%$, $+0.31\%$, and $+0.25\%$ NE respectively. The remaining modules together account for $+0.40\%$ NE in total. This suggests that, at the tested scale, feature interaction and fusion matter more than how they are encoded. Prior work invests most of its depth into attention stacks that process sequences in isolation. Our results indicate that redistributing some of that depth toward repeated cross-modal integration can be more beneficial.

\subsection{Interleaving Blocks and Residual Connections}
Results from the composition ablation (Table~\ref{tab:ablation_composition}) further support the above observation. When the same set of functional components is arranged sequentially rather than in a blocked, interleaved fashion, NE degrades by roughly $0.20\%$. Since the two configurations share identical components and comparable parameter counts, this gap is attributable entirely to the arrangement of components. As discussed in Section~\ref{sec:residual_flow}, the interleaved layout enables information to flow across module boundaries through residual connections. This creates cross-module information pathways that would require more explicit architectural additions in a sequential design. 
The $0.25\%$ NE contribution of the cross-module residual flow is notable because it adds zero parameters. In a sequential pipeline, residual connections within the attention stack carry only sequence information lacking context from cross-feature interaction stacks and vice versa. The stacked, interleaved design addresses this and transports information from both modalities to subsequent blocks where both attention and interaction can benefit from each other's contributions (Figure~\ref{fig:residual_flow}). The results of this ablation overall highlight the importance of layer and block composition which constitutes a core component of our contribution.

\subsection{Attention-Encoded User Interest}
Despite their lower overall contribution to the model's performance, the self and target attention modules both contribute meaningful signal to the integration stages which are the primary driver for \bbb's performance. The two encoding modules produce complementary interest signals from the input sequences, capturing global interest from self attention and candidate-specific interest from target attention. The integration stage can then jointly exploit these orthogonal signals. We attribute the relatively low individual contribution of each encoding component to its partial overlap in attended data which differs from prior approaches~\citep{zhai2024hstu} which use one type of attention as the sole sequence encoder.
\section{Limitations and Future Work}
While \bbb~demonstrates many strengths and consistent improvements over baseline architectures, we highlight several limitations that remain, which present opportunities for future work.

\subsection{Candidate-Aware Sequence Reweighting}
The sequence reweighting module conditions on full input sequences. While this design is effective in tuning similar sequences differently depending on user preference, it is agnostic to contextual implications of the current candidate items. This means that, independent of the candidate set, the weights produced by this step remain invariant within the same user. Incorporating candidate-aware information into the reweighting module could enable finer-grained personalization. Doing so safely and effectively poses several challenges. First, context must be injected without introducing leakage from sibling candidates. Second, processing must remain efficient. Na\"ive implementations that generate $\|C\|$ sequences for every candidate $c \in C$ are not feasible in production settings with sequence lengths in the thousands. A possible way to solve this could involve generating per-candidate residuals to modulate the centrally reweighted sequences.

\subsection{Block Assembly}
Our evaluation uses a fixed block composition of one full block followed by three long-sequence blocks. This configuration was derived empirically and validated with appropriate ablation studies (Section~\ref{sec:ablation}). However, the combinatorial potential of the presented components remains largely unexplored. To explore these more deeply, follow-up studies should investigate topics like the optimal ordering of block types, whether certain specializations should be placed earlier versus later in the stack, and how to balance full and lighter blocks for a given compute budget. Additionally, the presented block types (Section~\ref{sec:block_specializations}) are merely suggestions as to how blocks can be configured and represent only a subset of possible inter- and intra-block compositions.

\subsection{Throughput and Optimization Potential}
As shown in our evaluation, the interleaved design introduces a $5-10\%$ throughput overhead compared to the homogeneous attention stacks due to less favorable kernel fusion opportunities. Each block transitions between different operator types like attention, feature crossing, and reweighting, which limits the ability of current runtimes to fuse adjacent operations. Developing custom fused kernels for common sub-sequences within a \bbb~block could reduce this gap. More broadly speaking, this limitation highlights the importance of close co-design between kernel and model development. Future iterations could explore kernel and compiler-level optimizations to accelerate heterogeneous block designs like the one presented here.

\subsection{Scaling Approaches}
Although not the focus of the present study, we conduct preliminary scaling experiments with a simplified variant of \bbb~which we call \textit{BumbleBee Mini}. The model is configured with just the cross attention and feature interaction modules, without sequence personalization or self attention. 
We conduct experiments to explore two orthogonal scaling dimensions: (1) vertical scaling by adding more blocks at a fixed sequence length, and (2) trading off block depth against sequence length under a roughly fixed compute budget.

\subsubsection{Vertical Scaling}
We first evaluate the effect of increasing block depth at a fixed sequence length of 4096. As shown in Table~\ref{tab:scaling_depth}, adding layers yields consistent NE improvement: from two to three layers, NE improves by $-0.198\%$, and from three to four layers by an additional $-0.050\%$. While gains diminish with depth, they remain positive, suggesting that the interleaved design continues to extract value from additional blocks. A comprehensive study with the full \bbb~architecture---including self attention, target attention, and sequence reweighting---may exhibit more sustained scaling behavior, as each additional block would introduce richer information through these components.

\begin{table}[htbp]
\centering
\caption{Vertical scaling: increasing block depth at fixed sequence length (4096). NE improves consistently with additional layers, though with diminishing returns.}
\smallskip
\small
\begin{tabularx}{0.75\linewidth}{L R R}
\toprule
\textbf{Configuration} & \textbf{Mean NE (\%)} & \textbf{$\Delta$ vs.\ Baseline} \\
\midrule
4k seq, 2 layers (baseline) & 0.6109 & -- \\
4k seq, 3 layers & 0.6097 & $-0.198\%$ \\
4k seq, 4 layers & 0.6094 & $-0.248\%$ \\
\bottomrule
\end{tabularx}
\label{tab:scaling_depth}
\end{table}

\subsubsection{Sequence Length vs.\ Depth Trade-off}
We next explore whether longer input sequences can compensate for fewer blocks under a roughly constant compute budget. Table~\ref{tab:scaling_seqlen} shows that extending the sequence from 2048 to 4096 tokens while reducing depth from six to five layers improves NE by $-0.049\%$. Interestingly, further extending to 6144 tokens with four layers yields identical performance, suggesting a saturation point in the sequence length dimension at this scale. These results indicate that \bbb~supports flexible scaling along two orthogonal axes---depth and sequence length---and that practitioners can exploit the trade-off between the two to match their latency and memory constraints without sacrificing quality.

\begin{table}[htbp]
\centering
\caption{Sequence length vs.\ depth trade-off under roughly fixed compute. Longer sequences can compensate for reduced depth up to a saturation point.}
\smallskip
\small
\begin{tabularx}{0.75\linewidth}{L R R}
\toprule
\textbf{Configuration} & \textbf{Mean NE (\%)} & \textbf{$\Delta$ vs.\ Baseline} \\
\midrule
2k seq, 6 layers (baseline) & 0.6142 & -- \\
4k seq, 5 layers & 0.6139 & $-0.049\%$ \\
6k seq, 4 layers & 0.6139 & $-0.049\%$ \\
\bottomrule
\end{tabularx}
\label{tab:scaling_seqlen}
\end{table}

These initial results are promising and hint at the scalability of our design. Furthermore they show possible levers to trade-off performance based on how much data is available to the model. To better assess the full scaling potential of the design, a comprehensive study with the complete \bbb~architecture across a wider range of possible scaling approaches is needed to truly characterize the model's scaling behavior. Additionally, such a study will reveal whether a horizontal Mixture-of-Experts (MoE) composition can complement the vertical stacking approach presented here (Figure~\ref{fig:model_arch}).

\section{Conclusion}
In this paper, we introduce \bbb---a scalable, mixed-layer architecture for large-scale recommendation systems. Unlike prior approaches that process sequences and non-sequence features in separate, deep stacks, our design encourages early and repeated interaction between sequential and non-sequential features by interleaving sequence modeling and feature interaction components within self-contained blocks. Additionally, sequence personalization combined with target-aware encoding allows \bbb~to adapt to each user's interests more adequately than previous sequence modeling approaches. 

A key finding of our work is that the interleaved layout fundamentally changes the role of standard residual connections. These connections act as cross-module information pathways that transport \textit{both} sequence patterns and feature interaction signals between blocks. This mechanism contributes a $0.25\%$ NE improvement without adding any parameters, demonstrating that the strength of \bbb~lies less in individual component design and more in how components are composed. We support our design by a composition ablation which shows a $0.20\%$ normalized entropy (NE) improvement over a sequential arrangement of identical components.

The results of our evaluation and ablation study show that \bbb~outperforms state-of-the-art methods on industry scale recommendation tasks. The modular block design provides practical levers for trading off quality and throughput through careful block specialization. We believe the paradigm of interleaving heterogeneous functional units, rather than creating deep homogeneous stacks, offers a promising direction for creating more powerful and flexible building blocks for recommendation architectures beyond current designs.

\section{Acknowledgments}
\label{sec:acknowledge}
The authors wish to thank Yucheng Zhang, Eekin Chin, Sumedh Pendurkar, Fei Ding, Zhiyong Wang, David Cheung, Aowei Shen, Xiaoxing Zhu, Yasmine Badr, Hongbo Qin, Rui Zhang, Hong Li, Xuan Cao, Daisy Shi He, Kai Ren, Xinyao Hu, Shilin Ding and Meihong Wang for the collaboration and support. 

\clearpage
\newpage
\bibliographystyle{assets/plainnat}
\bibliography{main_document}



\end{document}